\documentclass[letters,fleqn,usenatbib]{mnras}

\usepackage{lmodern}
\usepackage{newtxtext,newtxmath}
\usepackage{float}
\usepackage[T1]{fontenc}
\usepackage{graphicx}
\usepackage{amsmath}
\usepackage{natbib}
\usepackage{url}
\usepackage{balance}
\usepackage{microtype}
\usepackage[dvipsnames]{xcolor}
\newcommand{\noopsort}[2]{#2}

\title[Nova super-remnant cavity surrounding RS\,Ophiuchi]{Discovery of a nova super-remnant cavity surrounding RS\,Ophiuchi}

\author[M. W. Healy-Kalesh et al.]{M. W. Healy-Kalesh,$^{1}$\thanks{E-mail: M.W.HealyKalesh@ljmu.ac.uk} M. J. Darnley,$^{1}$ \'E. J. Harvey,$^{1,2}$ and A. M. Newsam$^{1}$\\
$^{1}$Astrophysics Research Institute, Liverpool John Moores University, Liverpool, L3 5RF, UK \\
$^{2}$UK Astronomy Technology Centre, STFC, Royal Observatory, Edinburgh, EH9 3HJ, UK \\
}

\date{Accepted 2024 February 07. Received 2024 February 06; in original form 2023 September 05}

\pubyear{2024}

\begin{document}
\label{firstpage}
\pagerange{\pageref{firstpage}--\pageref{lastpage}}
\maketitle

\begin{abstract}
The prototypical nova super-remnant (NSR) was uncovered around the most rapidly recurring nova (RN), M31N\,2008-12a. Simulations of the growth of NSRs revealed that these large structures should exist around all novae, whether classical or recurrent. NSRs consist of large shell-like structures surrounding excavated cavities. Predictions, informed by these simulations, led to the discovery of an extended cavity coincident with the Galactic RN, RS\,Ophiuchi, in far-infrared archival IRAS images. We propose that this cavity is associated with RS\,Oph and is therefore evidence of another NSR to be uncovered.
\end{abstract}


\section{Introduction}\label{Introduction}
Recurrent novae (RNe) are a sub-type of cataclysmic variables that host an accreting white dwarf (WD) within a close binary system \citep{1954PASP...66..230W}. RNe  exhibit non-terminal thermonuclear driven eruptions \citep{1972ApJ...176..169S,2020ApJ...895...70S} where the inter-eruption time-scale is less than a century. During a nova eruption, much of the accreted material is ejected from the surface of the WD \citep{2020ApJ...895...70S} at velocities of hundreds to thousands of kilometers per second \citep{2001IAUS..205..260O}. This ejected material can lead to the formation of sub-parsec nova shells, \citep[see, e.g.,][]{2004ApJ...600L..63B,2007Natur.446..159S,2012ApJ...758..121S,2016A&A...595A..64H}. A proportion of the accreted material may remain to grow the WD \citep{2005ApJ...623..398Y,2007ApJ...659L.153H,2015ApJ...808...52K,2015MNRAS.446.1924H,2016ApJ...819..168H,2021gacv.workE..30S} toward the Chandrasekhar limit: potentially leading to a type Ia supernova \citep{1973ApJ...186.1007W,1999ApJ...519..314H,1999ApJ...522..487H,2000ARA&A..38..191H}.

A nova super-remnant (NSR) is a vast extended shell surrounding a RN formed by the cumulative effect of eruptions sweeping up local interstellar medium (ISM) over the lifetime of the system \citep{2019Natur.565..460D}. The first NSR found is located around the annually erupting RN, M\,31N 2008-12a \citep[12a; see][]{2016ApJ...833..149D,
2020AdSpR..66.1147D,2021gacv.workE..44D}; 12a is predicted to reach the Chandrasekhar limit in less than 20 kyr \citep{2017ApJ...849...96D}. Initially observed in 1992 within a H$\alpha$ survey \citep{1992A&AS...92..625W}, this NSR was `rediscovered' and associated with 12a in 2014 \citep{2015A&A...580A..45D}. The 12a NSR is substantially larger than any other nova shell, with semi-minor and -major axes extending to 90 and 134 parsecs, respectively \citep{2019Natur.565..460D}, rivalling the sizes of the largest known supernova remnants \citep{2001ApJ...563..816S}.

A preliminary study with one-dimensional hydrodynamical simulations \citep[using \texttt{Morpheus};][]{2007ApJ...665..654V} demonstrated the viability of NSR formation and persistence over the life-time of a nova system. \citet{2023MNRAS.521.3004H} presented an extensive suite of 1D hydrodynamical simulations exploring the impact of accretion rate, WD temperature and initial mass, and ISM density on NSR evolution. The \citet{2023MNRAS.521.3004H} study included the evolution of WD mass and the subsequent impact on ejecta properties, and incorporated radiative cooling. All \citet{2023MNRAS.521.3004H} NSRs consisted of a very low density excavated cavity surrounded by a hot ejecta pile-up region all contained within a thin, high-density shell of swept up ISM. 

\citet{2019Natur.565..460D} and \citet{2023MNRAS.521.3004H} demonstrated that dynamic NSRs should form around all nova systems, including classical novae \citep{2020AdSpR..66.1147D,2021gacv.workE..44D}. Furthermore, NSRs can exist around novae with growing {\it or eroding} WDs, indicating that NSRs should surround old novae with low-mass WDs  as well as RNe \citep{2023MNRAS.521.3004H}. Recently, a second NSR has been discovered around the Galactic RN, KT Eridani, through detection of its H$\alpha$ shell \citep{2023arXiv231017055S,2023arXiv231017258H}. A survey of M\,31 and LMC RNe did not find evidence for other NSRs \citep{2024arXiv240104583H}. Although such a dearth of NSRs may be due to the lower intrinsic luminosity of NSRs around systems with longer recurrence periods.

RS\,Ophiuchi is a Galactic symbiotic RN at a distance of $1.4^{+0.6}_{-0.2}$\,kpc \citep{2008ASPC..401...52B} or $2.4^{+0.3}_{-0.2}$\,kpc \citep{2021AJ....161..147B}, however this latter {\it Gaia} determination may be biased by the long-period RS\,Oph binary \citep{2018MNRAS.481.3033S}. RS\,Oph hosts a WD with mass $1.2-1.4\,\text{M}_{\odot}$ \citep{2017ApJ...847...99M} and a giant donor \citep[M0/2 III;][]{1994AJ....108.2259D,1999A&A...344..177A}. RS\,Oph has a substantial accretion rate to drive eruptions in 1898, 1933, 1958, 1967, 1985, 2006, and 2021 \citep{2022MNRAS.514.1557P}; exhibiting a recurrence period of ${\sim}15$ years \citep{2021gacv.workE..44D}. Eruptions in 1907 \citep{2004IAUC.8396....2S} and 1945 \citep{1993AAS...183.5503O} may have been missed due to Sun constraints \citep{2017ApJ...847...99M}.

Very Long Baseline Interferometry (VLBI) radio observations of RS\,Oph during its 1985 eruption revealed a bipolar structure at a position angle of $84^{\circ}$ \citep{1987rorn.conf..203P,1989MNRAS.237...81T}. This lack of spherical symmetry was also seen in VLBI and Multi-Element Radio-Linked Interferometer Network (MERLIN) radio images following the 2006 eruption with collimated bipolar jet-like flows expanding toward the East-West directions \citep{2006Natur.442..279O,2008ApJ...685L.137S,2008ApJ...688..559R,2013IAUS..281..195M}. Expanding lobes with the same East-West orientation were observed in late-time \textit{Hubble Space Telescope} and \textit{Chandra} X-ray imaging after the 2006 eruption \citep{2007ApJ...665L..63B,2009ApJ...703.1955R,2022ApJ...926..100M}. However, NIR and X-ray observations shortly after the 2006 eruption showed ejecta expanding in a different direction to the bipolar radio structure \citep{2007ApJ...658..520L,2009ApJ...707.1168L}.

As RS\,Oph has a short recurrence period, a massive WD, and a high accretion rate, it should be surrounded by a dynamic NSR grown over the life-time of the system \citep{2019Natur.565..460D,2020AdSpR..66.1147D,2021gacv.workE..44D,2023MNRAS.521.3004H}. In this Letter we present evidence for an excavated NSR cavity surrounding RS\,Oph. In Section~\ref{RS Ophiuchi NSR modelling} we predict the size of an RS\,Oph NSR; in Section~\ref{Finding a cavity around RS Oph} we describe the discovery of the RS\,Oph NSR cavity and its likely association with the nova system; the implications of uncovering another NSR are discussed in Section~\ref{Discussion}, before we present our conclusions in Section~\ref{Conclusions}.

\section{RS Ophiuchi NSR modelling}\label{RS Ophiuchi NSR modelling}

\subsection{RS Ophiuchi NSR dynamics}\label{RS Ophiuchi NSR dynamics}

\citet{2023MNRAS.521.3004H} presented a grid of NSRs  from novae with a range of accretion rates within a range of ISM densities. Here we utilise that grid to determine the dynamics of the NSR around an RS\,Oph-like system with a recurrence period of 15 years. 

\begin{figure}
\centering
\includegraphics[width=0.95\columnwidth]{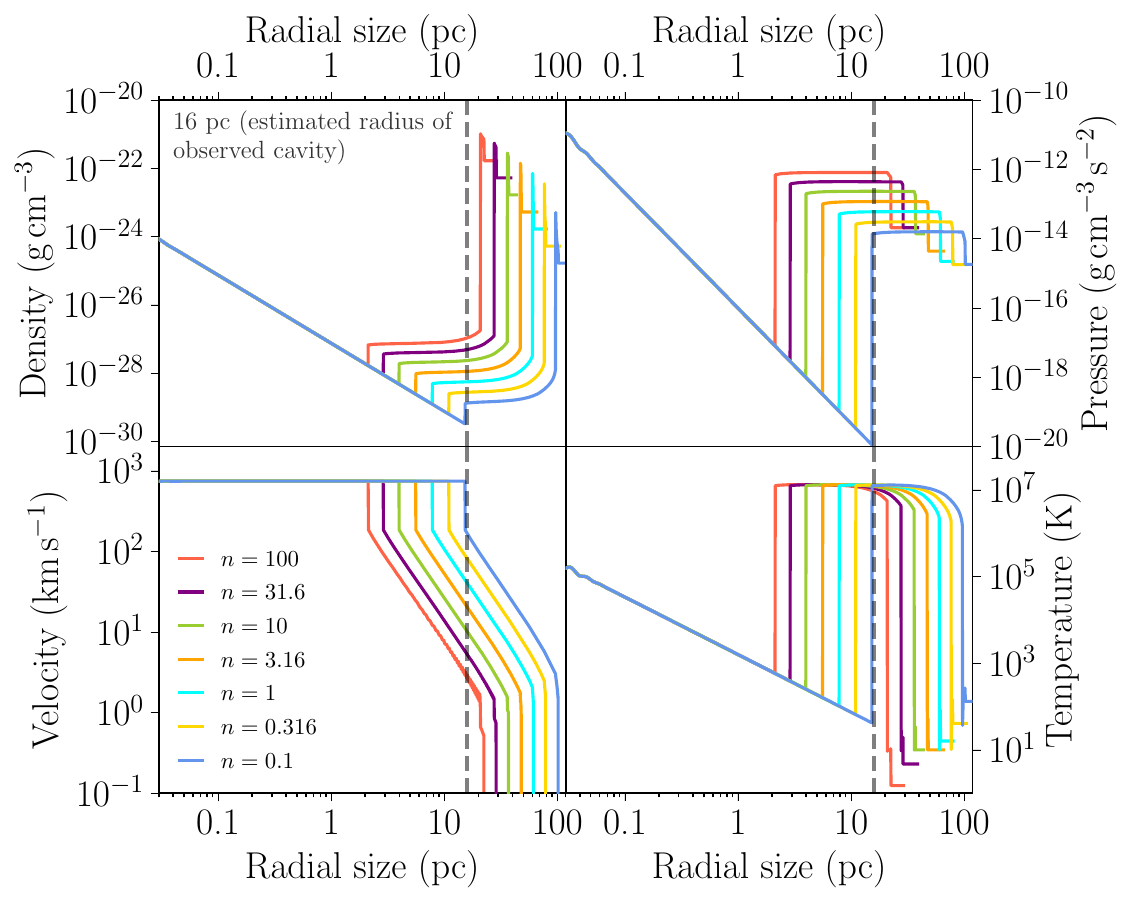}
\caption{Simulated dynamics of the NSR surrounding RS\,Oph from \citet{2023MNRAS.521.3004H} for different ISM densities.}
\label{RS Oph dynamics}
\end{figure}

We select models where a WD is grown from $1\,\text{M}_{\odot}$ with an accretion rate of $1\times 10^{-7}\,\text{M}_{\odot}\,\text{yr}^{-1}$ within ISM  ranging from $1.67\times10^{-25}\,\text{g}\,\text{cm}^{-3}$ ($n=0.1$) to $1.67\times10^{-22}\,\text{g}\,\text{cm}^{-3}$ \citep[$n=100$; Runs 1--7 from][]{2023MNRAS.521.3004H}. These models predict that a $1\,\text{M}_{\odot}$ WD takes ${\sim}25$\,Myr ($\sim$650,000 eruptions) to grow to $1.27\,\text{M}_{\odot}$, i.e., a recurrence period of 15 years. We show the density, pressure, velocity and temperature characteristics of the NSR at this stage in Figure~\ref{RS Oph dynamics} for a range of ISM densities.

As shown in Figure~\ref{RS Oph dynamics}, NSR radial size decreases with increasing ISM density. Specifically, the outer edges of the NSR shells (and corresponding shell thicknesses) extend to 102 (5\%), 79 (3\%), 62 (2\%), 48 (2\%), 37 (3\%), 29 (4\%) and 22 pc (7\%) for $n=10^{-1}, 10^{-0.5}, 10^0, 10^{0.5}, 10^1, 10^{1.5}, 10^2$, respectively. The velocity of the outer shell is $\lesssim3\,\text{km}\,\text{s}^{-1}$ for all NSRs and the temperature at the outer edges range from 130\,K ($n=0.1$) to 2\,K ($n=100$). The size of the NSR cavity, the inner-most region, also decreases with increasing ISM density. The largest simulated cavity, ${\sim}$16\,pc, occurs for $n=0.1$. The cavity density is predicted to be up to four orders of magnitude lower than the surrounding ISM.

\subsection{ISM density and NSR cavity prediction}\label{RS Ophiuchi NSR size prediction}
To estimate the ISM density in the region around RS\,Oph we took two approaches. The first  utilises the Besan\c{c}on model of the Galaxy\footnote{\url{http://cdsarc.u-strasbg.fr/viz-bin/qcat?J/A+A/453/635}} \citep[see, for e.g.,][]{2006A&A...453..635M}, which provides the distribution of $K$-band extinction along different lines of sight. The second method uses the three-dimensional dusts maps\footnote{\url{http://argonaut.skymaps.info}} from \citet{2019ApJ...887...93G}. In both cases, we used the Galactic coordinates and distance to RS\,Oph to find the relevant grid cell containing the system. Optical extinction\footnote{$K$-band extinction from the Besan\c{c}on model was converted to $V$-band extinction using \citet{1989ApJ...345..245C}.} was converted to hydrogen column density using the relation in \citet{2009MNRAS.400.2050G}, assuming Solar abundances \citep[$X = 0.739$;][]{2004ApJ...606L..85B}. The two methods yielded $n_1 = 0.032 \pm 0.001$ and $n_2 = 0.066 \pm 0.003$, respectively. While statistically inconsistent, neither method is favoured, therefore we assume an ISM density around RS\,Oph of $n=0.05\pm0.02$.

As the \citet{2023MNRAS.521.3004H} grid doesn't include ISM as sparse as $n=0.05$, we select the $n=0.1$ model as the most suitable for predicting the current size of the RS\,Oph NSR shell and cavity (see Figure~\ref{RS Oph dynamics}). As such, we predict that the radius of any NSR shell surrounding RS\,Oph should be $\gtrsim100$\,pc, with the radius of the inner evacuated cavity extending to $\gtrsim15$\,pc. With RS\,Oph at a distance of $1.4^{+0.6}_{-0.2}$\,kpc, this equates to an angular size of $\gtrsim250^\prime$ and $\gtrsim40^\prime$ for the diameters of the NSR shell and cavity, respectively. Following the methodology outlined in \citet{2023MNRAS.521.3004H}, we predict the total X-ray luminosity and H$\alpha$ luminosity from the RS\,Oph NSR to be ${\lesssim}10^{25}$ erg s$^{-1}$ and ${\lesssim}4 \times 10^{31}$ erg s$^{-1}$, respectively.

\section{The search for a cavity around RS Ophiuchi}\label{Finding a cavity around RS Oph}

\subsection{Structure in IRIS data}\label{Structure in IRIS data}

Given the approximate size of a potential RS\,Oph NSR, a search for a structure of similar dimensions was conducted in the vicinity. We concluded that the signature of a NSR cavity might be particular evident in the far-IR due to the `missing' cool gas. The search therefore included the new generation {\it Infrared Astronomical Satellite} ({\it IRAS}) catalogue, known as IRIS \citep[Improved Reprocessing of the {\it IRAS} Survey; ][]{2005ApJS..157..302M}.

\begin{figure*}
\centering
\includegraphics[width=0.4\textwidth]{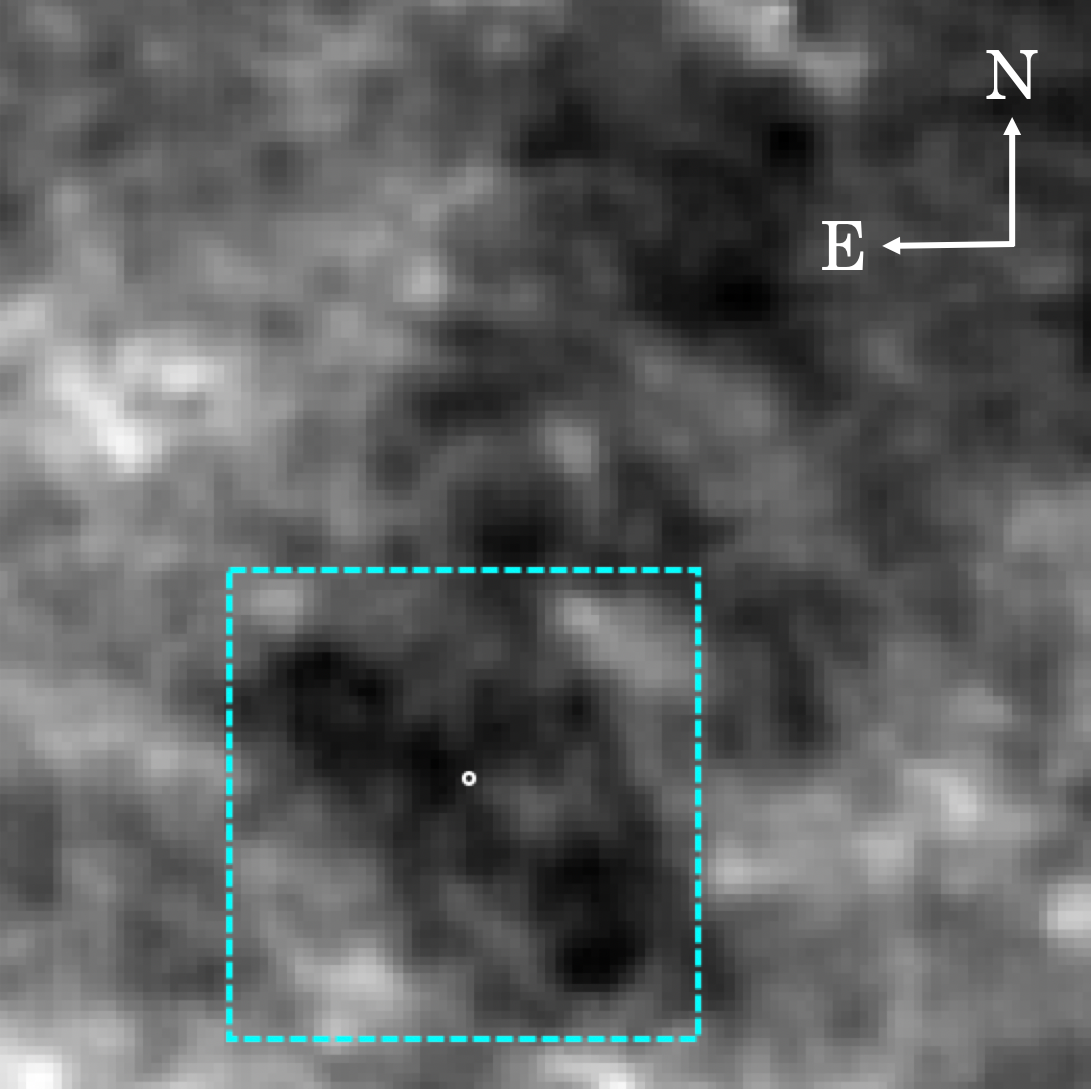}
\label{IRIS full image}
\centering
\includegraphics[width=0.4\textwidth]{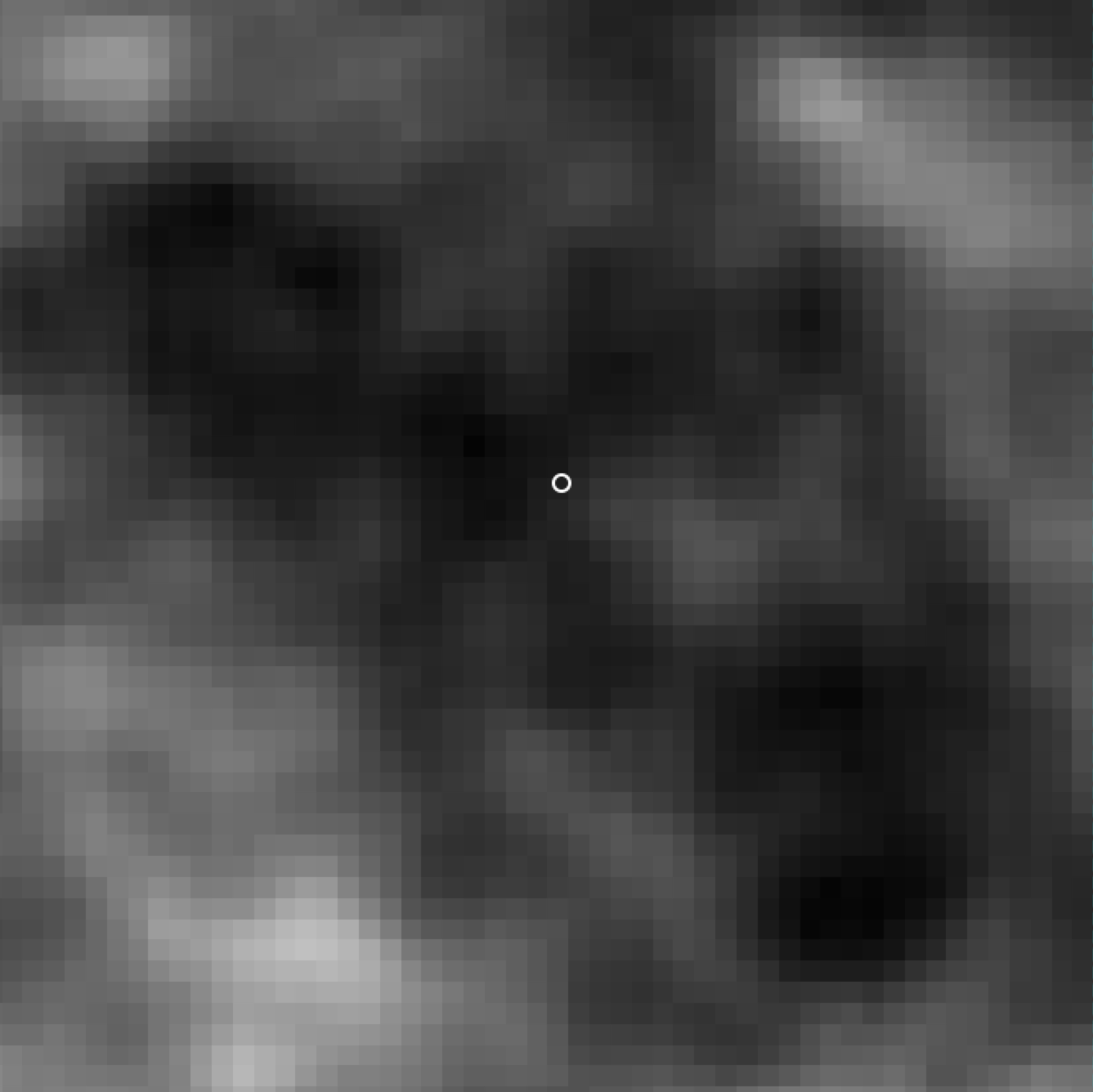}
\label{IRIS zoom image}
\caption{Cavity surrounding RS\,Oph. The white circles are the current location of RS\,Oph. {\bf Left}: A $3^{\circ} \times 3^{\circ}$ IRIS image of the surroundings of RS\,Oph in the far infrared (100\,$\mu$m). {\bf Right}: A $1.3^{\circ} \times 1.3^{\circ}$ region from the left image (see cyan box).}
\label{RS Oph cavity}
\end{figure*}

A cavity-like structure was promptly found within a $3^{\circ} \times 3^{\circ}$ IRIS 100\,$\mu$m image containing RS\,Oph (see Figure~\ref{RS Oph cavity}). The cavity has semi-major and -minor axes of approximately ${\sim}40^{\prime}$ and ${\sim}12^{\prime}$, with a position angle of ${\sim}50^\circ$. At a distance of ${\sim}1.4$ kpc \citep{2008ASPC..401...52B}, this equates to a physical size of ${\sim}16 \times 5$\,pc. While only ${\sim}6\%$ of the size of the predicted overall NSR, this is a close match to the predicted extent of NSR cavities simulated by \citet{2023MNRAS.521.3004H}. The sheer scale of this cavity might suggest a reason why this structure around RS Oph has not be associated with the RN until now.

\subsection{Determining association between cavity and RS Oph}\label{Determining association between cavity and RS Oph}

To confidentially associate this far-IR cavity with RS\,Oph, we must exclude a coincidental alignment between two unrelated phenomena. Here we define a cavity to be any elliptical region where the flux per unit sky area is less than the flux per unit sky area within a surrounding elliptical annulus. We denote the ratio between the fluxes of the annulus and the cavity to be $\mathscr{R}$; $\mathscr{R}>1$ defines a cavity.

\subsubsection{Reference ellipse}\label{Reference ellipse}

For a reference ellipse, the ellipticity and position angle are free parameters; the position is fixed to the location of RS Oph and the semi-major axis is fixed at ${\sim}40^\prime$ (a projected size of ${\sim}16$\,pc), to match the prediction of the cavity size from simulations. The inner and outer semi-major-axes of the annulus are defined as 1.0 and 1.5 times the semi-major axis of the cavity. 

\begin{figure*}
\centering
\includegraphics[width=\textwidth]{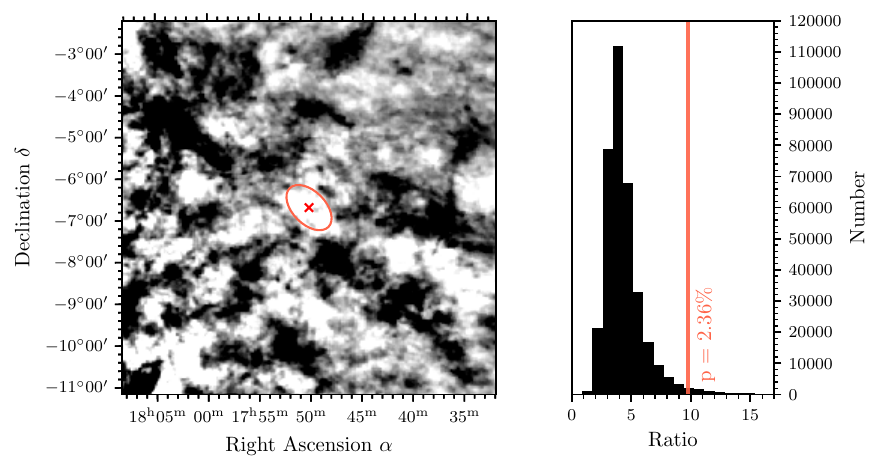}
\caption{{\bf Left}: An inverted $9^{\circ} \times 9^{\circ}$ IRIS 100\,$\mu$m centred on RS\,Oph (red cross). The best fitting elliptical cavity $\mathscr{E}_{\text{cavity}}$ (see text) is illustrated by the red ellipse. {\bf Right}: Histogram illustrating the distribution of cavity ratios $\mathscr{R}$ computed at 361,201 locations within the IRIS data (left). The horizontal red line denotes $\mathscr{E}_{\text{cavity}}$.}
\label{E_cavity and histogram}
\end{figure*}

With this reference ellipse cavity fixed at the position of RS\,Oph, we find that an ellipticity of 0.4 and a position angle of 45$^{\circ}$ produces the largest ratio $\mathscr{R} = 9.74$. This best fitting cavity is illustrated in Figure~\ref{E_cavity and histogram} (red ellipse) and thus defines our border for the cavity around RS\,Oph. We refer to this best-fitting ellipse as $\mathscr{E}_{\text{cavity}}$.

\subsubsection{Monte Carlo analysis}\label{Monte Carlo analysis}

\begin{figure}
\centering
\includegraphics[width=\columnwidth]{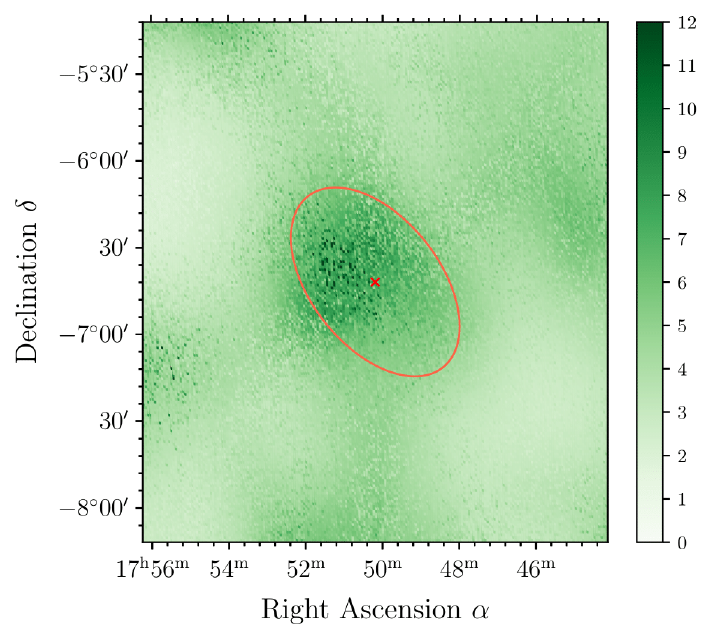}
\caption{Heat map of ellipse  ratio values $\mathscr{R}$ for the ellipse $\mathscr{E}_{\text{cavity}}$ as a function of position within the IRIS data (see Figure~\ref{E_cavity and histogram}). The elliptical cavity $\mathscr{E}_{\text{cavity}}$ is shown centred on the position of RS\,Oph.}
\label{Heat map}
\end{figure}

A Monte Carlo analysis utilised a $9^{\circ} \times 9^{\circ}$ background subtracted IRIS image centred on RS\,Oph. We sampled $\sim$360,000 ellipses, each with a randomly selected semi-major axis, ellipticity and position angle on a uniformly distributed spatial grid across the image and computed $\mathscr{R}$ at each position. The semi-major axis is taken from a range of radial sizes (${\sim}16-22$ pc) derived from linearly interpolating between $n = 0.05 \pm 0.02$ to match estimates of ISM density around RS Oph (see Section~\ref{RS Ophiuchi NSR size prediction}); ellipticity is taken from a range derived from the system having an inclination between $0-70^{\circ}$, and the position angle is taken from a range between $0-180^{\circ}$. The resultant distribution of the computed $\mathscr{R}$ for all the ellipses is shown in the right panel of Figure~\ref{E_cavity and histogram} and as a heat map in Figure~\ref{Heat map}.

We find that 97.64\% of the sample of ellipses (without any fixed parameters) have a ratio $\mathscr{R} < 9.74$. As such, the probability of a chance alignment of RS\,Oph with a cavity that has less far-IR emission than the $\mathscr{E}_{\text{cavity}}$ cavity is $2.36\%$. We note that Figure~\ref{Heat map} demonstrated that many positions with $\mathscr{R} > 9.74$ are in a small region near to the position of RS\,Oph; which further strengthens the association between the cavity and RS\,Oph. As $\mathscr{R}$ is determined by both the cavity and the surrounding emission, the heat map may not be a reliable source of the position of the cavity centre.

\section{Discussion}\label{Discussion}

Having identified a large structure centred on the location of the RN RS\,Oph where we expect to find a vast NSR, we have explored the possibility of coincidental alignment. We found that the probability of finding a cavity-like structure such as this in the same vicinity ($9^{\circ} \times 9^{\circ}$) of the sky is $<2.36\%$ and so the location of this cavity is significant.

We would expect there to be various structures within the Galactic medium with similar properties to the proposed NSR cavity surrounding RS\,Oph. For example, superbubbles are cavities created within the ISM by the winds from OB associations \citep{1986PASJ...38..697T,1988srim.conf..447M,2020NewAR..9001549W}. From inspection of the most well-studied OB associations provided in \citet{2020NewAR..9001549W}, we find that there is no superbubble spatially coincident with RS\,Oph \citep[the closest being Ser OB2 with an angular radius of 31 pc;][]{2010MNRAS.402.2369T,2017MNRAS.472.3887M} so we rule out this as a possibility.

Other structures that can span similar areas on the sky are supernova remnants (SNRs). Cross-matching the large sample of 294 SNRs in \citet{2019JApA...40...36G} with the location of our discovered RS Oph cavity reveals zero overlapping remnants as they are all confined to the Galactic plane, as do the 132 SNRs provided in the Chandra Supernova Remnant Catalog\footnote{\url{https://hea-www.cfa.harvard.edu/ChandraSNR/snrcat_gal.html}} \citep[a proportion of these SNRs are the same as those in][]{2019JApA...40...36G}. A search for other sources within an approximately 20 arcminute radius of RS Oph using the SIMBAD database \citep{2000A&AS..143....9W} displays a number of X-ray sources, including a compact object \citep{2012ApJ...756...27L}, and one radio source, however none of the identified sources could be responsible for the found cavity.

We can also consider the orientation of RS\,Oph which has a large impact on the shaping and geometry of the shells within its immediate vicinity. The position angle of the found cavity (${\sim}45^{\circ}$) does not match the value derived from the resolved shell around RS\,Oph created during the 2006 eruption \citep[$85^{\circ}$;][]{2009ApJ...703.1955R}. Yet, the cavity does resemble the general shape of a much more extended and evolved bipolar structure \citep{2006Natur.442..279O,2009ApJ...703.1955R} that we would expect to grow over the lifetime of RS\,Oph as a RN if it were consistently producing `dumbbell-shaped' remnants \citep{2009ApJ...703.1955R}. Furthermore, the likely inhomogeneous material surrounding RS\,Oph at much greater distances than the immediate vicinity may contribute to the discrepancy between position angle of the cavity and the shell grown from a single eruption.

\begin{figure}
\centering
\includegraphics[width=0.9\columnwidth]{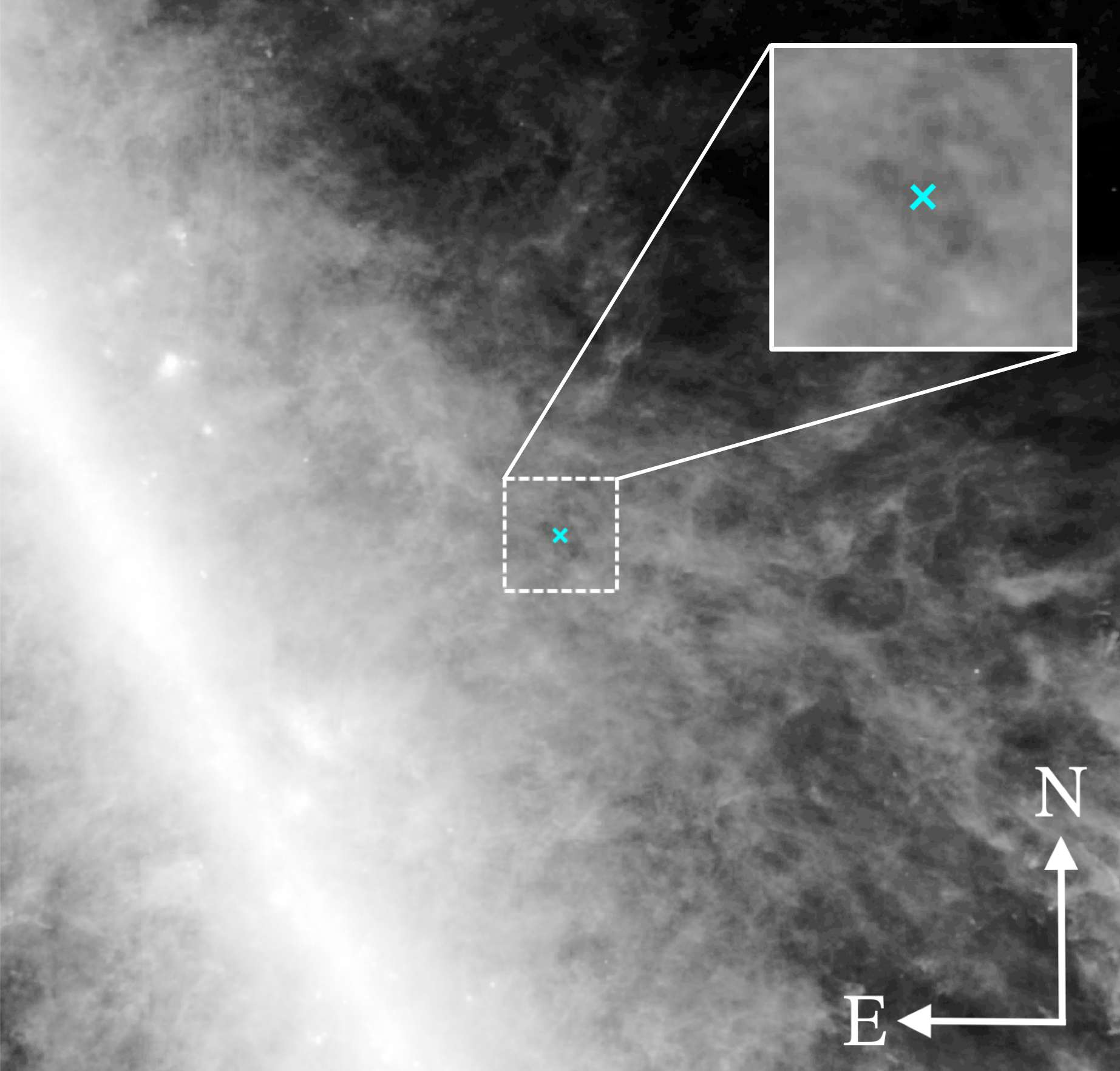}
\caption{Cavity surrounding RS\,Ophiuchi can be identified in a $30^{\circ} \times 30^{\circ}$ IRAS image. The cyan cross indicates the location of RS\,Oph. The white dashed-line box is a $3^{\circ} \times 3^{\circ}$ region which we show in the inset.}
\label{fig:30 by 30}
\end{figure}

Though we have outlined (and subsequently ruled out) a number of alternative explanations for the cavity, the most compelling finding we present for its association with RS\,Oph is the elliptical cavity we have identified is situated with its major axis running through the location of RS\,Oph, with a very low probability of a coincident alignment. In fact, the cavity structure possibly associated with RS\,Oph can be seen in a $30^{\circ} \times 30^{\circ}$ far-infrared image as shown in Figure~\ref{fig:30 by 30}. We note that there also appears to be a ring-like structure to the northwest of the elliptical cavity which may have also been shaped by RS\,Oph. However, we can not speculate further on this.

Uncovering a further example of a NSR through the detection of its cavity component around another RN, RS Oph, reinforces the hypothesis that these structures are created from many prior nova eruptions sweeping up the local ISM \citep{2019Natur.565..460D,2023MNRAS.521.3004H}. Additionally, finding this likely NSR cavity within far-IR data encourages broadening our search for more NSRs to infrared wavelengths alongside H$\alpha$ and X-ray emission.

\section{Conclusions}\label{Conclusions}

The RN M31N\,2008-12a hosts the prototypical nova super-remnant: an immense dynamically-grown shell with a major axis greater than 100 parsecs formed from the continual sweeping of local ISM. Similar NSRs are predicted to surround other novae as it is the eruptions emanating from the central WD that drives their growth. 

In far-infrared IRAS imaging, we have uncovered the cavity of such a NSR coincident with the Galactic RN RS\,Oph and we find that it is very likely that they are associated. Narrow-band imaging of the surroundings of RS Oph akin to those which were employed to detect the NSR shell associated with KT Eri \citep{2023arXiv231017055S,2023arXiv231017258H}, would cement this connection. Ultimately, if this cavity is part of a more extended NSR, then this finding strengthens the hypothesis that all RNe should be surrounded by a NSR and further indicates that the 12a NSR is not unique.

\section*{Acknowledgements}
The authors would like to gratefully thank our anonymous referee for their time spent reviewing our manuscript and for their constructive feedback that improved our study. MWH-K, MJD, and {\'E}JH recieved funding from the UK Science and Technology Facilities Council grant  ST/S505559/1. The authors thank Prof.\ Steven Longmore for assistance estimating local ISM density. This work made use of the high performance computing facilities at Liverpool John Moores University, partly funded by LJMU’s Faculty of Engineering and Technology and by the Royal Society. This research has made use of the SIMBAD database, operated at CDS, Strasbourg, France \citep[see][]{2000A&AS..143....9W}. This research has made use of the VizieR catalogue access tool, CDS, Strasbourg, France. The original description of the VizieR service was published in \citet{2000A&AS..143...23O}.

\section*{Data Availability}
Data in this study can be shared on reasonable request to the corresponding author. The analysis in this work made use of the Python libraries: Numpy \citep{harris2020array} and Matplotlib \citep{Hunter:2007}.
\balance

\bibliographystyle{mnras}
\bibliography{bibliography}

\bsp
\label{lastpage}
\end{document}